\documentstyle[epsfig]{l-aa}

\begin{document}

\def\ltsima{$\; \buildrel < \over \sim \;$}
\def\simlt{\lower.5ex\hbox{\ltsima}}

   \thesaurus{11.19.1 -- 13.25.2}
   \title{Dust lanes, thick absorbers, and the unification
model for Seyfert galaxies}

   \author{Giorgio Matt
          }


   \institute{Dipartimento di Fisica ``E. Amaldi",
              Universit\`a degli Studi Roma Tre,
              Via della Vasca Navale 84, I--00146 Roma, Italy
	}

   \date{Received / Accepted }

   \maketitle

\markboth{Giorgio Matt: Dust lanes, thick absorbers, and the unification
model for Seyfert galaxies}{}

   \begin{abstract}
A modification of the popular unification model for Seyfert galaxies 
is proposed, which takes into account recent observational findings on the 
statistical properties of both type 1 and type 2 Seyferts. 

In the proposed scenario, Compton--thick Seyfert 2 galaxies are those sources
observed through compact, thick matter (the `torus'), while Compton--thin/
intermediate Seyfert galaxies are obscured by dust lanes at larger distances.
\keywords{Galaxies: Seyfert -- X-rays: galaxies }
   \end{abstract}

%

\section{Introduction}

The discovery of broad lines in the polarized flux of the archetypal
Seyfert 2 galaxy NGC~1068 (Antonucci \& Miller 1985) has been a 
landmark in the study of AGN, leading to the now widely accepted 
unification model for Seyfert galaxies (e.g. Antonucci 1993). In this
scenario, type 1 and type 2 Seyferts are intrinsically the same, 
appearing different only because in the former we can see the nucleus
directly, while in the latter the direct view 
is prevented by absorbing matter on the line--of--sight. 

There can be no doubt about the basic correctness of the unification model.
In fact, there is plenty of examples of Seyfert 2s with unambiguously
have an obscured type 1 nucleus at their centre, while we are not aware of 
even a single Seyfert 2 which certainly
does not harbour a hidden Seyfert 1. However, it is likely that the strictest
version of the model (in which the aspect angle is the only relevant
parameter) is not fully valid. Arguments against it include:

\begin{itemize}
\item a) there is, on average, enhanced star formation in Seyfert 2 galaxies 
with respect to Seyfert 1s (Maiolino et al. 1997); 

\item b) the average morphologies between galaxies hosting type 1 and 2 nuclei are 
different, those hosting
type 2 being on average more irregular (Maiolino et al. 1997, Malkan et al.
1998);

\item c) there is a greater overall dust content in Seyfert 2s (Malkan et al. 1998). 
\end{itemize}

\begin{figure*}[t]
\centerline{\epsfig{figure=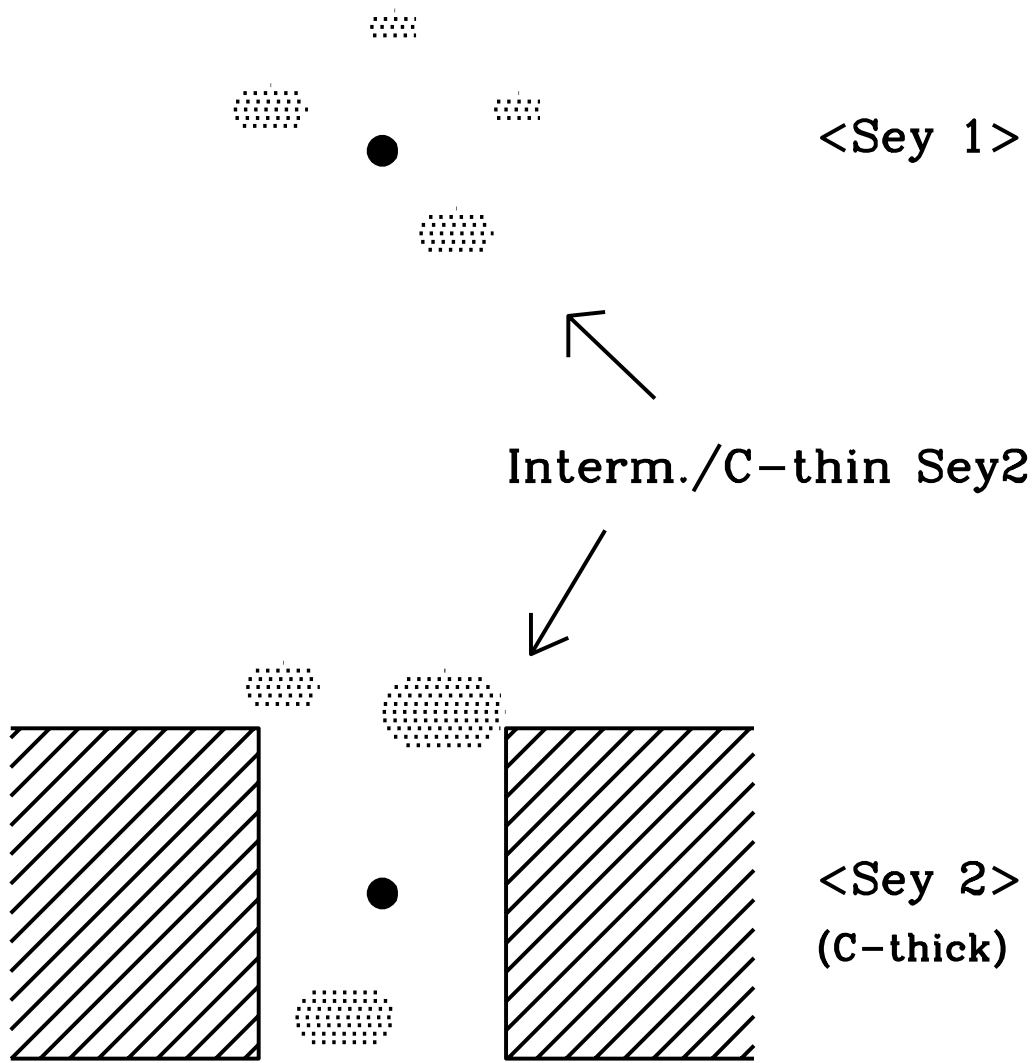,height=18.0cm, width=18.0cm}}
\caption{The proposed unification model (see text for details).}
\label{umplot}
\end{figure*}

Moreover, Malkan et al. (1998) showed that 
there is plenty of dust lanes at distances of hundred of parsecs
in all type of Seyfert galaxies. They 
went as far as to propose that these dust lanes are completely responsible for 
the type 1/type 2 dichotomy, and
therefore dismissing the existence of the torus.
To avoid confusion, it is important to remark that
here and after we use the term `torus' to indicate any distribution
of optically thick matter close (a few tens of parsecs at most) to the 
nucleus, and with a large covering factor, whatever its actual geometry is
(not necessarily ring--shaped!).
Actually, there are many good arguments in favour of the
existence of the torus. Apart from those listed by Antonucci (1993),
more recent ones include radio imaging and water maser measurements, 
indicating dense
matter very close to the black hole (e.g. Gallimore, Baum \& O'Dea 1997;
Greenhill et al. 1996); and infrared imaging of nearby Seyfert 2s,  again
indicating the presence of large amount of matter very close to the nucleus 
(e.g. Siebenmorgen et al. 1997).

In the last few years, very strong evidence in favour of the `torus' 
(whatever it really is) has been obtained from X--rays observations. 
In particular, BeppoSAX observations have shown 
that at least half of Seyfert 2s in the local Universe are Compton--thick
(Maiolino et al. 1998a; Risaliti, Maiolino \& Salvati 1999), i.e. the nuclear
radiation is absorbed 
by matter with column densities exceeding 10$^{24}$ cm$^{-2}$ 
(see Matt et al. 2000 for the general 
properties of bright Compton--thick Seyfert 2s). While in a handful of
Compton--thick sources the column density has been directly measured
(e.g. NGC~4945: Iwasawa et al. 1993, Done et al. 1996, Guainazzi et al. 2000;
Circinus Galaxy: Matt et al. 1999; NGC~6240: Vignati et al. 1999), in the 
majority of them either the column density is so large to completely obscure
the nucleus even in hard X--rays (e.g. NGC~1068: Matt et al. 1997) or their
flux at high energies is too low to permit a detailed spectral analysis or,
often, 
even a detection with the present generation of hard X--ray detectors. 
In this case, the classification of a source as Compton--thick 
lies on indirect arguments: a reflection--dominated spectrum (recognized
by the flat slope, if the reflector is `cold', and by a $\sim$1 keV equivalent
width iron line) is the most useful and used indicator.

The observed large fraction of Compton--thick Seyfert 2s 
implies that the covering factor of such thick matter 
must be large. Assuming a spherical geometry for simplicity,
 the total amount of matter is proportional to the square of the outer radius,
provided that it is much larger than the inner radius (this
argument holds, at least roughly, whatever is the geometry, if the 
covering factor is large). In order not to exceed the value of the mass
obtained from dynamical measurements, the outer radius of the torus in
Circinus Galaxy must be less than 20 pc (Maiolino et al. 1998b). A less
tight constraint is derived from NGC~1068 (Risaliti et al. 1999), i.e
$\simlt$100 pc, which however still 
implies that the dust lanes on the hundred parsecs
scale  cannot be the matter responsible for the absorption in this source.

A further important finding of Risaliti et al. (1999) is that there is
a clear difference between the $N_H$ distribution of Intermediate 
(type 1.8--1.9) and strict type 2 Seyferts. While the intermediate Seyferts
in the Risaliti et al. sample are all Compton--thin, the strict type 2
Seyferts have column densities generally exceeding 10$^{23}$ cm$^{-2}$, and
most of them are Compton--thick. 

In the following section we will discuss a possible modification of the 
unification model which qualitatively accounts for
the different statistical properties of obscured and unobscured Seyfert
galaxies, and for the different column density distribution of intermediate
and strict type 2 sources. 

\section{A modification of the unification model}

The proposed modification of the unification model is illustrated in Fig.~1.
The basic properties are as follows:

\begin{itemize}

\item In all Seyferts there are dust lanes on scales of hundred of
parsecs, as observed by Malkan et al. (1998).
These lanes have column densities 
of the order of 10$^{22}$--10$^{23}$ cm$^{-2}$
at most, otherwise the mass involved would be too large. 
The fact that the dust content of Seyfert 2s appears, on average,
to be greater than that
of Seyfert 1s may be related to the more disturbed morphology of the Seyfert 2s
host galaxies, possibly as a result or a recent interaction with another
galaxy.

\item Not all Seyferts have the torus (or, at least, not all have
a torus with a large covering factor). Again, there may be a greater
chance of producing a torus in Seyfert 2s, as they are more disturbed and
with a larger overall dust content. 

\end{itemize}

There are, therefore, three different possibilities:

\begin{itemize}

\item The sources observed through a dust lane (but outside the torus) are
the Compton--thin (in X--ray terminology) or intermediate (in optical 
terminology) Seyferts. 

\item The sources observed through the torus are the strict Seyfert 2s
(most of them Compton--thick, using the X--ray terminology).
 
\item If the line--of--sight to the nucleus is free of any
absorber, the source is a Seyfert 1. Of course, it is more likely
(but not necessary) that a source is observed as Seyfert 1 when 
the torus is not present. 

\end{itemize}

It is worth remarking that 
the fraction of sources with the thick torus must be fairly large, as
Compton--thick sources account for at least half of the total number
of obscured Seyferts (Risaliti et al. 1999).

\section{Discussion}

The proposed  modification, 
 while retaining the basic and most successful characteristic 
of the unification model (i.e. that all intermediate
and type 2 Seyferts harbour an obscured type 1 nucleus), explains, at least
qualitatively, the observed 
differences between the average properties
of Seyfert 1s and Seyfert 2s and between the column density distributions
of intermediate and strict type 2 Seyferts. This model is somewhat 
different (even if on a similar line of thought)
from that proposed by Maiolino \& Rieke (1995), and takes into
account recent observational results. The 
two `flavours' of Seyferts, with and without the thick torus, may represent 
either two different branches in the AGN evolution, or a different evolutionary
stage in the life of any (or most) source. 

Testing the proposed model would require further studies of the statistical
properties of Seyferts, which caution in separating intermediate and
strict type 2 sources. For instance, as the torus is expected to be
axially symmetric (while the dust lanes are probably more randomly distributed),
a correlation between Compton-thick absorption and presence of ionization cones
and large polarization is expected. Another test is to 
search for the presence of the torus in Seyfert 1s, as we predict that many 
sources of this class do not have it.  This may be done either by 
searching for strong IR emission (which however may be related to starburst
rather than reprocessing of UV/X--rays from thick matter), of by searching
for signatures of X--ray reprocessing (e.g. Ghisellini, Haardt \& Matt 1994;
Krolik, Madau \& \.Zycki 1994). The latter way have already produced a clear
case of a Seyfert 1 {\it with} the torus: NGC~4051 was caught by BeppoSAX when
the nuclear emission was switched--off, and clear evidence for reprocessing
by large amount of optically thick distant matter was present 
(e.g. a $\sim$600 eV equivalent width iron line and a cold reflection
continuum, Guainazzi et al. 1998). 
However, this kind of
observations based on variability requires rather extreme behaviours of 
the X--ray emission, and it is not clear how common a switching--off
of the nucleus can be. More promising is to search for narrow (i.e. unresolved)
iron K$\alpha$ lines in addition to the relativistically broadened component
from the accretion disc. When the nuclear X--ray emission is directly visible, 
equivalent widths of the order of 50--100 eV 
are expected from the torus  (Ghisellini, Haardt \& Matt 1994). While ASCA
and BeppoSAX results have been rather ambiguous in this respect, 
such a search is certainly within the capabilities of XMM.

\begin{acknowledgements}
I acknowledge financial support from ASI and MURST (grant
{\sc cofin}98--02--32).
\end{acknowledgements}


\begin{thebibliography}{} 

\bibitem[]{}  Antonucci R., 1993, ARA\&A 31, 473

\bibitem[]{}  Antonucci R.R.J. and Miller J.S., 1985, ApJ 297, 621

\bibitem{} Done C., Madjeski G.M., Smith D.A., 1996, ApJ 463, L63


\bibitem[]{} Gallimore J.F., Baum S.A., O'Dea C.P., 1997, Nat, 388, 852 

\bibitem[]{} Ghisellini G., Haardt F., Matt G., 1994, MNRAS 267, 743

\bibitem[]{} Greenhill, L. J., Gwinn, C. R., Antonucci, R., Barvainis, R.,
1996, ApJ 472, L21

\bibitem[]{} Guainazzi M., Nicastro F., Fiore F., et al., 1998, MNRAS, 301, L1

\bibitem[]{} Guainazzi M., Matt G., Brandt W.N., et al., 2000, A\&A, in press

\bibitem[]{} Krolik J.H., Madau P., \.Zycki P.T., 1994, ApJ, 420, L57

\bibitem{} Iwasawa K., Koyama K., Awaki H., et al., 1993, ApJ 409, 155

\bibitem[]{} Maiolino R., Rieke G.H., 1995, ApJ, 454, 95

\bibitem[]{} Maiolino R., Ruiz M., Rieke G.H., Papadopoulos P., 1997,
        ApJ, 485, 552

\bibitem[]{} Maiolino R., Salvati M., Bassani L., et al., 1998a, A\&A, 338, 781

\bibitem[]{} Maiolino R., Krabbe A., Thatte N., Genzel R., 1998b, ApJ, 493, 650

\bibitem[]{}  Malkan M.A., Gorjian V., Tam R., 1998, ApJS, 117, 25

\bibitem{} Matt G., Guainazzi M., Frontera F., et al., 1997, A\&A, 325, L13

\bibitem{} Matt G., Guainazzi M., Maiolino R., et al., 1999, A\&A 341, L39

\bibitem[]{} Matt G., Fabian A.C., Guainazzi M., et al., 2000, MNRAS, submitted

\bibitem[]{} Risaliti G., Maiolino R., Salvati M., 1999, ApJ, 522, 157

\bibitem[]{} Siebenmorgen R., Moorwood A., Freudling W., K\"{a}ufl H. U., 1997,
A\&A, 325, 450

\bibitem{} Vignati P., Molendi S., Matt G., et al., 1999, A\&A, 349, L57

\end{thebibliography}
\end{document}